\documentclass{revtex4-1}
\usepackage{color,graphicx,epsfig} 
\usepackage{ifpdf}
\usepackage{amsmath} 
\usepackage{amssymb} 
\usepackage{bm}
\usepackage{color}

\pdfoutput=1

\usepackage[english]{babel}
\usepackage{slashed}
\usepackage{multirow}

\begin{document}

\author{Ilja Dor\v sner} \email[Electronic address:]{ilja.dorsner@ijs.si}
\affiliation{Department of Physics, University of Sarajevo, Zmaja od Bosne 33-35, 71000
  Sarajevo, Bosnia and Herzegovina}

\author{Svjetlana Fajfer} \email[Electronic
address:]{svjetlana.fajfer@ijs.si} 
\affiliation{Department of Physics,
  University of Ljubljana, Jadranska 19, 1000 Ljubljana, Slovenia}
\affiliation{J.\ Stefan Institute, Jamova 39, P. O. Box 3000, 1001
  Ljubljana, Slovenia}

\author{Ivana Musta\' c}
\email[Electronic address:]{ivana.mustac@ijs.si}
\affiliation{J.\ Stefan Institute, Jamova 39, P. O. Box 3000, 1001 Ljubljana, Slovenia}

\title{Light vector-like fermions in a minimal $SU(5)$ setup}

\date{\today}

\begin{abstract}
The Standard Model fermion sector is enlarged by either one light singlet vector-like down-type quark or one light vector-like lepton doublet, which might be accommodated within a five-dimensional representation of $SU(5)$. At low energies the inclusion of these states affects precisely measured observables in flavor physics, as well as electroweak precision measurements. These experimental results strongly constrain couplings of vector-like states to the Standard Model particles. Having these bounds, we investigate the impact of vector-like fermions on the mass matrices for down-type quarks and charged leptons in an $SU(5)$ setting. We find that unitary transformations relating an arbitrary flavor basis to the mass eigenstate basis depend only on three free parameters. Then we discuss the parameter space constrained by low-energy data assuming  vector-like quark and vector-like lepton masses to be 800\,GeV and 400\,GeV, respectively. We demonstrate that these two scenarios generate unique patterns for relevant proton decay widths. A further improvement of experimental bounds on proton decay modes would thus differentiate the allowed parameter space. We finally present two full-fledged $SU(5)$ models that allow for gauge coupling unification with light vector-like fermions under consideration and discuss their viability.
\end{abstract}

\maketitle


\section{Introduction}
\label{INTRODUCTION}

The LHC discovery of the Higgs boson has finally confirmed the correctness of the Standard Model (SM) picture of fundamental interactions~\cite{Aad:2012tfa,Chatrchyan:2012ufa}. There is, however, a number of open issues that still point towards potentially new physics. These issues concern, for example, the origin of neutrino masses, the nature of dark matter, a hierarchy problem, the question of vacuum stability and the so-called flavor puzzle. One class of proposals, among many, that aims to address some of these issues, calls for the presence of vector-like fermions. These have been introduced either as a part of effective theories~\cite{Dermisek:2012as,Joglekar:2012vc} or within more elaborate frameworks such as Little Higgs models~\cite{Han:2003wu,Perelstein:2003wd}, composite Higgs models~\cite{Contino:2006qr,Anastasiou:2009rv,Matsedonskyi:2012ym} and grand unified theory (GUT) models~\cite{Berezhiani:1985in,Davidson:1987mi,Hisano:1993uk,Babu:1995fp,Berezhiani:1995yk,Babu:1995hr,Babu:1995uu,Shafi:1999ft,Barr:2003zx,Malinsky:2007qy,Oshimo:2009ia,Babu:2012pb,Barr:2012ma,Barr:2013dca}. 

Vector-like fermions are primarily introduced in GUT models to modify mass relations among quarks and leptons~\cite{Witten:1979nr}. Namely, the original $SU(5)$ model~\cite{Georgi:1974sy} predicts masses of down-type quarks and charged leptons to be degenerate at the scale of gauge coupling unification, i.e., the GUT scale $m_\mathrm{GUT}$. However, the running of these masses to low energies yields substantially different values from experimentally observed ones (for a review see Ref.~\cite{Babu:2009fd}). There are different ways to correct these erroneous mass predictions. One possibility is to add extra scalar multiplet(s) within $SU(5)$~\cite{Georgi:1979df}. The main drawback of this approach is the loss of predictive power due to the fact that these extra scalar multiplets have large dimensions~\cite{Giveon:1991zm,Dorsner:2006dj,Dorsner:2007fy,Dorsner:2009mq,Dorsner:2011ai,Dorsner:2013tla}. Another possibility is the introduction of higher-dimensional operators in the Yukawa sector of the theory. Again, the price to pay is the loss of predictive power due to the presence of many possible terms. Needless to say, both of these possible modifications spoil the simplicity of the original setup. A third approach is to add vector-like fermions to the theory. This is the direction we plan to pursue. 

We add one vector-like down-type quark and one vector-like lepton doublet that comprise a pair of five-dimensional representations of $SU(5)$ in order to obtain correct values for the masses of down-type quarks and charged leptons. A number of studies to include such fermions already exists for the $SU(5)$ GUT models with~\cite{Shafi:1999ft,Barr:2003zx,Oshimo:2009ia,Babu:2012pb} and without supersymmetry (SUSY)~\cite{Barr:2012ma,Barr:2013dca}. However, all of the existing studies take the masses of the vector-like states to be at the GUT scale. In our work, on the other hand, we assume vector-like fermions to be light enough to be accessible at LHC and study current experimental constraints on the couplings of these states to the matter fields. Recent phenomenological analyses of vector-like fermions in different representations can be found for example in \cite{Fajfer:2013wca,Dawson:2012di,Aguilar-Saavedra:2013qpa,Cacciapaglia:2010vn,Cacciapaglia:2011fx,Altmannshofer:2013zba,Falkowski:2013jya,Arhrib:2009mz}. The minimal vector-like fermion extension of the original $SU(5)$ proposal requires very few new couplings, preserves renormalizability and reproduces mass relations among charged leptons and down-type quarks in accordance with experimental data. Our study demonstrates that the proposed scenario is very predictive with regard to proton decay signatures through both scalar and gauge boson mediation. We furthermore present two full-fledged $SU(5)$ models that yield gauge coupling unification, accommodate all experimentally observed fermion masses, and allow for light vector-like states we study.

Light vector-like fermions necessarily affect low-energy observables. However, the low-energy flavor phenomena involving quarks are rather well described within the SM framework. This, therefore, strongly restricts the effective couplings of vector-like quarks with the matter fields and the gauge bosons (see~\cite{Fajfer:2013wca} and references therein). Direct searches at LHC have also produced bounds on the masses of vector-like quarks~\cite{Chatrchyan:2012af,B2G-12-019,B2G-12-021,B2G-13-003,ATLAS-CONF-2013-056}. The only dedicated study that provides direct bounds on the masses of vector-like leptons we consider is provided by the LEP L3 collaboration~\cite{Achard:2001qw}. The effective couplings of these states to the SM fields are constrained through experimental data on processes such as $\mu$--$e$ conversion in nuclei and lepton number violating decays. 
 
Our work is organized as follows. In Section~\ref{CONSTRAINTS} we present current constraints coming from the low-energy phenomenology on the presence of a vector-like quark that is an isosinglet with charge $-1/3$. That section also contains a discussion of constraints on the presence of a light vector-like charged lepton. We allow these new vector-like quarks (leptons) to mix with all three generations of matter fields in order to have the most general scenario. Section~\ref{FRAMEWORK} contains the study of the impact the vector-like quarks and leptons, accommodated in one pair of five-dimensional representations of $SU(5)$, have on the mass relations between the down-type quarks and charged leptons.
Section~\ref{PREDICTIONS} is devoted to consequences of the presented scenario on two-body proton decay due to gauge boson exchange after we impose the relevant low-energy constraints. To demonstrate the viability of the proposed vector-like extension we present two full-fledged $SU(5)$ models in Section~\ref{MODELS}. We conclude briefly in Section~\ref{CONCLUSIONS}.


\section{Constraints}
\label{CONSTRAINTS}

The new degrees of freedom we plan to introduce in the $SU(5)$ framework include one leptonic vector-like isodoublet and one down-type quark vector-like isosinglet. Both of these states can couple to the SM fermions through Yukawa and Dirac interactions, affecting the phenomenology of the electroweak sector. We parametrize the consequences from low-energy flavor and electroweak precision observables in terms of matrices in flavor space that originate from an additional mismatch between gauge and mass basis, triggered by the presence of these new fermions. In this framework, we assume the vector-like matter to be the lightest new degrees of freedom, and to consequently present the leading effects on low-energy interactions. We study two limiting cases, namely the one in which we investigate effects coming solely from the additional quark-like states assuming the vector-like leptons to be heavy enough in order to be neglected in the calculations, and vice versa.

\subsection{Vector-like quarks}
\label{vld}

The presence of Dirac and Yukawa interaction terms mixing SM and vector-like states requires an additional diagonalization of the quark mass matrices, on top of the standard Cabbibo-Kobayashi-Maskawa (CKM) mechanism, being performed through unitary rotations in the left-  and right-handed sector.
Since standard and vector-like representations differ in one of their chiralities, tree level flavor changing neutral interactions are generated, as opposed to the flavor conserving neutral currents of the SM. In this case, i.e., when an isosinglet vector-like representation is added, the leading effects occur among the left-handed states. The modified interactions among the Z boson and the SM down-type quarks, in the mass basis, can be written as
\begin{equation}
\label{eq:LZq}
\mathcal L_Z \supset - \frac{g}{c_W} \left( - \frac{1}{2} X^d_{ij} \bar d^i \gamma^\mu P_L d^j + \frac{1}{3} s_W^2 \bar d^i \gamma^\mu d^i  \right) Z_\mu\,,
\end{equation}
with $i,j= 1, 2, 3$. Here $P_{L,R} = (1 \pm \gamma_5)/2$, $g =2m_W/v \simeq 0.65$ is the weak coupling, while $s_W \equiv \sin\theta_W = \sqrt{0.231}$ and $c_W = \sqrt{1-s_W^2}$ are the sine and cosine of the weak angle, respectively. The second part contains the SM electromagnetic quark current, while the matrix $X^d$ ($X^d_{ii} \equiv 1- \delta X^d_{ii}$) incorporates the diagonal SM part as well as the new physics contribution. The elements of $X^d$ are directly connected to the unitary transformations, relevant for the GUT framework phenomenology, and the processes which affect these parameters will give the leading constraints, whereas charged current interactions give milder bounds. (See, for example, Ref.~\cite{Fajfer:2013wca}.) Note that flavor changing neutral currents (FCNCs) appear solely in the down-type quark sector in this setup.
Strong bounds on $\delta X^d_{ii}$ come from $Z$-pole physics, while the off-diagonal interactions are severely restricted by measurements on mesonic FCNC decays. The numerical upper limits on the entries of $X^d$, obtained in \cite{Fajfer:2013wca}, are displayed in Table~\ref{Zcoups} and are subsequently used in Section~\ref{FRAMEWORK}.

\begin{table}
\begin{centering}
\begin{tabular}{|c|c|}
\hline
Coupling & Constraint \tabularnewline
\hline
\hline
${\rm }|X^d_{12}|$ & $ 1.4\times 10^{-5}$ \tabularnewline
\hline
$|X^d_{13}|$ & $ 0.4 \times 10^{-3}$ \tabularnewline
\hline
$|X^d_{23}|$ & $ 1.0 \times 10^{-3}$ \tabularnewline
\hline
\hline
$\delta  X_{11}^{d}$ & $ 4.0 \times 10^{-3}$ \tabularnewline
\hline
$\delta  X_{22}^{d}$ & $ 6.0 \times 10^{-3}$ \tabularnewline
\hline
$\delta  X_{33}^{d}$ & $ 5.7 \times 10^{-3}$ \tabularnewline
\hline
\end{tabular}
\par\end{centering}
\caption{\label{Zcoups}Phenomenological upper bounds on $Z$ couplings to the SM quarks (see Eq.~\eqref{eq:LZq}) from precision flavor and electroweak observables. All upper bounds are given at 95\% C.L..}
\end{table}


\subsection{Vector-like leptons}

Contrary to the previous case, where we have introduced a new isosinglet, the addition of an isodoublet vector-like state manifests itself primarily in the right-handed fermion sector. Since we do not consider the issue of neutrino masses, i.e., we do not add any field that would allow for the generation of neutrino mass terms, there are no effects on charged interactions among the SM leptons. The modified neutral interactions between the SM charged leptons can be written as
\begin{equation}
\mathcal L_Z \supset - \frac{g}{c_W} \left( - Y^e_{ij} \bar e^i \gamma^\mu P_R e^j - \frac{1}{2} \bar e^i \gamma^\mu P_L e^i + s_W^2 \bar e^i \gamma^\mu e^i  \right) Z_\mu\, . \label{LZl}
\end{equation}
As noted before, in the fit of non-standard lepton couplings we did not include $X^d$. The constraints on the diagonal entries of $Y^e$ coming from $Z$-pole physics, \cite{ALEPH:2005ab}, \cite{Altmannshofer:2013zba}, are given in Table~\ref{lepZ}.
\begin{table}
\begin{centering}
\begin{tabular}{|c|c|}
\hline
Coupling & Constraint \tabularnewline
\hline
\hline
${\rm }|Y^e_{12}|$ & $ 1.6 \times 10^{-7}$ \tabularnewline
\hline
$|Y^e_{13}|$ & $ 5.5 \times 10^{-4}$ \tabularnewline
\hline
$|Y^e_{23}|$ & $ 5.5 \times 10^{-4}$ \tabularnewline
\hline
\hline
$Y^e_{11}$ & $ 6.8 \times 10^{-4}$ \tabularnewline
\hline
$Y^e_{22}$ & $ 2.8 \times 10^{-3}$  \tabularnewline
\hline
$Y^e_{33}$ & $ 2.0 \times 10^{-3}$ \tabularnewline
\hline
\end{tabular}
\par\end{centering}
\caption{\label{lepZ} Phenomenological upper bounds on $Z$ couplings to SM leptons (see Eq.~\eqref{LZl}) from precision flavor and electroweak observables, where $Y^e_{ij}$ is the new physics contribution to $Z$ couplings in the right-handed sector. All upper bounds are given at 95\% C.L.\ and the theoretical constraint $Y^e_{ii} > 0$ was taken into account.}
\end{table}

The $\mu$--$e$ conversion in nuclei sets a very severe bound on $Y^e_{12}$, since it occurs at tree level in this setup. One can write the branching ratio in the form
\begin{equation}
\mathcal{B}_{\mu N \to e N} = \frac{8 G_F^2}{\omega_{\rm cap.}}\ | Y^e_{12}|^2\ |(2 g_u + g_d) V^{(p)} + (g_u + 2 g_d) V^{(n)}|^2 \, ,
\end{equation}
\cite{Kitano:2002mt}, \cite{Altmannshofer:2013zba}. The muon capture rate  $\omega_{\rm cap.}$ and the overlap integrals $V^{(p)}$ and $V^{(n)}$ can be found in \cite{Kitano:2002mt} and
\begin{equation}
g_u = 1 - \frac{8}{3} s_W^2\ , \qquad g_d = -1 + \frac{4}{3} s_W^2
\end{equation}
embody the vector couplings of up-type and down-type quarks to the $Z$ boson, respectively. Using the existing experimental upper limits on $\mu$--$e$ conversion in gold and titanium atoms shown in Table~\ref{lepbrs}, we get $|Y^e_{12}| < 1.6 \times 10^{-7}$ at 95\% C.L.. This is the limit we present in Table~\ref{lepZ}.

\begin{table}
\begin{centering}
\begin{tabular}{|c|c|}
\hline
Observable & Constraint \tabularnewline
\hline
\hline
$\mathcal{B}_{\mu {\rm Au} \to e {\rm Au}}$ & $ 0.7 \times 10^{-12}$ \cite{Bertl:2006up}  \tabularnewline
\hline
$\mathcal{B}_{\mu {\rm Ti} \to e {\rm Ti}}$ & $ 1.7 \times 10^{-12}$ \cite{Kaulard:1998rb}  \tabularnewline
\hline
$\mathcal{B}(\tau \to 3e)$ & $ 2.7 \times 10^{-8}$ \cite{Hayasaka:2010np}  \tabularnewline
\hline
$\mathcal{B}(\tau \to 3\mu)$ & $ 2.1 \times 10^{-8}$ \cite{Hayasaka:2010np} \tabularnewline
\hline
\end{tabular}
\par\end{centering}
\caption{\label{lepbrs} Experimental upper limits on several lepton flavor violating processes at 90\% C.L..}
\end{table}

Lepton flavor violating decays of the form $l_i \to 3l_j$ give a weaker bound in the $\mu$--$e$ sector, while they deliver the leading constraint on mixed couplings involving the $\tau$ lepton in our model. At tree level in this setup, the corresponding width can be written as \cite{Kamenik:2009cb}
\begin{equation}
\Gamma(\tau \to l_i l_i \bar{l_i}) = \frac{G_F^2}{48 \pi^3} m_{\tau}^5\ |Y_{i3}^e|^2 \left[ \left(s_W^2 -\frac{1}{2} \right)^2 + \frac{3}{2} \left(s_W^2 - Y_{ii}^e \right)^2 \right]\, ,
\end{equation}
with $l_i =e$ or $\mu$. The result in Table~\ref{lepZ} was obtained by marginalizing over the ranges of $Y^e_{ii}$ allowed by $Z$-pole physics. All constraints that are summarized in Table~\ref{lepZ} are used in Section~\ref{FRAMEWORK}.

Finally, future experiments will explore a large portion of the presently allowed parameter space, since the sensitivity to the mentioned processes is supposed to improve by various orders of magnitude. The PRISM/PRIME~\cite{Barlow:2011zza} and furthermore the Mu2e~\cite{Abrams:2012er} experiments, for example, are expected to probe $\mu$--$e$ conversion rates to the order of $10^{-18}$ and $10^{-17}$, respectively. A dedicated experiment for measuring $\mu \to 3e$ is planned to reach a sensitivity of $10^{-16}$~\cite{Blondel:2013ia}, while an improvement in sensitivity by about an order of magnitude is foreseen by SuperKEKB regarding $\tau \to 3l$~\cite{Aushev:2010bq}. Moreover, several other proposals for new experiments or upgrades in the area of precision physics will affect a wide range of additional processes, which will potentially make them more relevant than the mentioned ones. At this place a comment should be made on $\mu \to e \gamma$. Since also in this model it occurs only through a loop, it is highly suppressed compared to $\mu$--$e$ conversion in nuclei, both processes being directly proportional to $|Y^e_{12}|^2$. At present, the experimental bounds on the branching ratios are of the same order of magnitude, $\mathcal{B}(\mu \to e \gamma) < 5.7 \times 10^{-13}$~\cite{Adam:2013mnn} (compare with the entries for $\mu$--$e$ conversion in nuclei in Table \ref{lepbrs}), making the tree level process more sensitive to the newly induced coupling. Finally, the MEG experiment might improve the present limit on $\mu \to e \gamma$ by one order of magnitude~\cite{Baldini:2013ke}.


\subsection{Collider phenomenology}

The ATLAS and CMS collaborations at the LHC have published several dedicated studies on the direct production of vector-like quarks, with the result of lower bounds on the particles' masses, due to the non-observation of such states. These searches focus on the QCD pair production of the exotic states with subsequent decay into third generation SM quarks and gauge bosons or the Higgs. The latest of these analyses combine the possibilities of decays into the mentioned three channels, $Q \to V q$, with $V=W,Z,h$, and $q$ a top or a bottom quark. In the large mass limit, for masses $\gtrsim$ 500\,GeV, typically half of the vector-like quarks decay through the charged current channel, while the branching ratios in each of the neutral channels amount to $\sim 25\%$, which are usually referred to as the nominal branching fractions. 
The most recent CMS studies report a lower bound of around 700\,GeV for the nominal branching fractions~\cite{B2G-12-019},~\cite{B2G-13-003}, while a recent ATLAS study delivers a limit of 645\,GeV \cite{ATLAS-CONF-2013-056} for that case, all at 95\% C.L.. The mentioned searches do not include couplings to first two generation quarks. Nevertheless, at the masses probed by now, this generalization would not alter the outcome significantly in the most minimal models, since low energy processes highly restrict the couplings to lighter quarks. The latter, on the other hand, becomes extremely relevant in the case of single production. This might provide a relevant channel for future studies~\cite{Buchkremer:2013bha}, since the production rate overcomes that of single production at higher masses.

A general direct search for exotic leptons has in turn been performed at LEP by the L3 collaboration \cite{Achard:2001qw}, setting a lower bound on the mass of charged leptons, which are part of a vector-like isodoublet, at about 100\,GeV. Moreover, studies by the LHC experiments searching for heavy leptons in the context of the Type III see-saw model have been made, with ATLAS setting the most stringent bound of 245\,GeV~\cite{ATLAS-CONF-2013-019}. Since the decays of the exotic states depend on their mixing with SM leptons, these were varied in the study, however without considering mixing with the $\tau$. In~\cite{Redi:2013pga} it was shown that in a composite scenario with leptonic vector-like isodoublets, this bound is lifted to about 300\,GeV. The authors of Ref.~\cite{Falkowski:2013jya} performed a recast of a CMS multilepton search using the full dataset at $\sqrt{s}=8$ TeV \cite{SUS-13-002}. Assuming couplings to electrons or muons only, they obtained a bound of about 460\,GeV, while in the case of mixing with the $\tau$ alone they computed a weaker limit of about 280\,GeV. However, it should be noted that in the mentioned work, carried out in a composite Higgs framework, there is no contribution from decays of exotic to the SM leptons and the Higgs boson.


\section{$SU(5)$ setup}
\label{FRAMEWORK}

The $SU(5)$ setup we study comprises matter fields that belong to $\bm{10}_i=\{e^C_i,u^C_i,Q_i\}$ and $\overline{\bm{5}}_{i}=\{L_i,d^C_i\}$, $i=1,2,3$, where $Q_i=(u_i
\quad d_i)^T$ and $L_i=(\nu_i \quad e_i)^T$~\cite{Georgi:1974sy}. It also contains one vector-like pair $(\bm{5}_4,\overline{\bm{5}}_4)$ of matter fields, where $\bm{5}_4 =\{\overline{L}_4 ,\overline{d}^C_4 \}$ and $\overline{\bm{5}}_4 =\{L_4 ,d^C_4 \}$. The subscript for the vector-like pair is included to allow for more compact notation. This pair will be used to generate viable masses for down-type quarks and charged leptons. 

The scalar sector of the setup, on the other hand, is made out of one $24$-dimensional and one $5$-dimensional representation we denote as $\bm{24}$ and $\bm{5}$, respectively. The adjoint representation $\bm{24}$ breaks $SU(5)$ down to $SU(3) \times SU(2) \times U(1)$ while the fundamental representation $\bm{5}$ provides the electroweak vacuum expectation value (VEV). We take $|\langle \bm{5}^5 \rangle| = v'$ and $\langle \bm{24} \rangle \approx \sigma \,\mathrm{diag} (2,2,2,-3,-3)$
to be the relevant VEVs, where $\bm{5}^\alpha$, $\alpha=1,\ldots,5$, represent components of the fundamental representation of $SU(5)$. We neglect the contribution of an $SU(2)$ triplet towards the VEV of $\bm{24}$. The exact value of $\sigma$ can be determined through the consideration of gauge coupling unification. It is constrained to be of the same order as the scale at which gauge couplings meet, i.e., the GUT scale, due to experimental input on proton decay. The common value of the SM gauge couplings at $m_\mathrm{GUT}$ is $g_\mathrm{GUT}$.

The up-type quark masses originate from a single $SU(5)$ operator $(Y^{10})_{ij} \bm{10}_i \bm{10}_j \bm{5}$, $i,j=1,2,3$, where $Y^{10}$ is a complex $3 \times 3$ Yukawa matrix. The contraction in the space of flavor is explicitly shown for clarity. The important thing to note is that the up-type quark mass matrix comes out to be symmetric. This feature will be relevant for the proton decay predictions we present in Section~\ref{PREDICTIONS}. 

The down-type quark and the charged lepton masses, on the other hand, require a more elaborate structure in order to be viable. Namely, one requires the presence of two types of operators to generate realistic masses. These are $(Y^{\overline{5}})_{il}  \bm{10}_i \overline{\bm{5}}_{l} \overline{\bm{5}}$ and $\overline{\bm{5}}_{l} [ (M)_l+(\eta)_l \bm{24}] \bm{5}_4$, where $i=1,2,3$ and $l=1,2,3,4$. Here, $M$ is an arbitrary complex $1 \times 4$ mass matrix while $\eta$ and $Y^{\overline{5}}$ represent Yukawa matrices with complex entries of dimensions $1 \times 4$ and $3 \times 4$, respectively. It is the latter set of operators that breaks the degeneracy of the mass spectrum of vector-like leptons and vector-like quarks within the $(\bm{5}_4, \overline{\bm{5}}_4)$ pair.

We can redefine matter field multiplets at the $SU(5)$ level to go to the basis where the term $(Y^{\overline{5}})_{i4}  \bm{10}_i \overline{\bm{5}}_{4} \overline{\bm{5}}$ is completely removed and the remaining $3 \times 3$ part of the $Y^{\overline{5}}$ matrix with components $(Y^{\overline{5}})_{ij}$, $i,j=1,2,3$, is diagonal~\cite{Babu:2012pb}, i.e., $(Y^{\overline{5}})_{ij}=y^{\overline{5}}_i \delta_{ij}$. In this basis the $4 \times 4$ mass matrices $M_E$ and $M_D$ that are relevant for the charged lepton and the down-type quark sectors explicitly read 
\begin{equation}
\label{4times4}
M_E = \left( \begin{array}{cccc}
y^{\overline{5}}_1 v'   & 0   & 0 & M^E_1 \\
0 & y^{\overline{5}}_2 v' & 0 & M^E_2 \\
0 & 0 & y^{\overline{5}}_3 v' & M^E_3 \\
0 & 0 & 0 & |M^E_4|
\end{array} \right),\qquad
M_D = \left( \begin{array}{cccc}
y^{\overline{5}}_1 v'   & 0   & 0 & 0 \\
0 & y^{\overline{5}}_2 v' & 0 & 0 \\
0 & 0 & y^{\overline{5}}_3 v' & 0 \\
M^D_1 & M^D_2 & M^D_3 & |M^D_4|
\end{array} \right),
\end{equation}
where $M^{E}_l=(M)_l-3(\eta)_l \sigma$ and $M^{D}_l=(M)_l+2(\eta)_l \sigma$, $l=1,2,3,4$. Again, we neglect the contribution from the VEV of the $SU(2)$ triplet in $\bm{24}$ towards the mass of the vector-like leptons. Our convention is such that $M^D$ is multiplied from the left by a $1 \times 4$ matrix $(d_1\quad d_2\quad d_3\quad \overline{d}^C_4)$ and from the right by a $4 \times 1$ matrix $(d^C_1\quad d^C_2\quad d^C_3\quad d^C_4)^T$.

It is possible to make all matrix elements in Eq.~\eqref{4times4} real by making suitable redefinitions of the quark and lepton fields~\cite{Babu:2012pb}. We will assume that this is done and neglect in the rest of our work these phases for simplicity. With this in mind we introduce the parameters $m_i=|y^{\overline{5}}_i| v'$ and $x_i^{E,D}=M^{E,D}_i/|M^{E,D}_4|$, $i=1,2,3$. These parameters will play a crucial role in our study of fermion masses and mixing parameters. We will use this basis and associated nomenclature as the starting point for our discussion. Note that the CKM phase will come exclusively from the up-type quark sector in this framework.

If one omits contributions from the $(\bm{5}_4,\overline{\bm{5}}_4)$ pair, one finds that the down quark mass $m_1^D$ and the electron mass $m_1^E$ are both equal to $m_1$ at the GUT scale and thus degenerate. This is in disagreement with experimental observations, once the measured masses are propagated from the low-energy scale to $m_\mathrm{GUT}$. In fact, the same type of degeneracy would also hold for the masses of the second and the third generation of down-type quarks and charged leptons, i.e., $m_j^D=m_j^E$, $j=2,3$. This, again, does not correspond to what experimental values yield at $m_\mathrm{GUT}$. It is the presence of vector-like matter that breaks this degeneracy and creates an opportunity to have a realistic scenario. 

It can be shown~\cite{Babu:2012pb} that the charged fermion masses $m_i^E$, $i=1,2,3$, are related to the parameters $m_i$ and $x^{E}_i$ through the following three equations
\begin{equation}
\label{eq:1}
(m_1^E)^2+(m_2^E)^2+(m_3^E)^2=\frac{m_1^2(1+|x^E_2|^2+|x^E_3|^2)+m_2^2(1+|x^E_3|^2+|x^E_1|^2)+m_3^2(1+|x^E_1|^2+|x^E_2|^2)}{1+|x^E|^2},
\end{equation}

\begin{equation}
\label{eq:2}
(m_1^E m_2^E)^2+(m_1^E m_3^E)^2+(m_2^E m_3^E)^2=\frac{m_1^2 m_2^2(1+|x^E_3|^2)+m_2^2 m_3^2(1+|x^E_1|^2)+m_3^2 m_1^2(1+|x^E_2|^2)}{1+|x^E|^2},
\end{equation}

\begin{equation}
\label{eq:3}
(m_1^E m_2^E m_3^E)^2=\frac{m_1^2 m_2^2 m_3^2}{1+|x^E|^2},
\end{equation}
where we introduce $|x^E|^2=|x_1^E|^2+|x_2^E|^2+|x_3^E|^2$. These relations should be satisfied at the GUT scale.

The important point is that these equations are also applicable for the down-type quark sector. All one needs to do is to replace $m_i^E$ with $m_i^D$ and $x^{E}_i$ with $x^{D}_i$ in Eqs.~\eqref{eq:1},~\eqref{eq:2} and~\eqref{eq:3}. The parameters $m_i$, on the other hand, are common for both sectors. Clearly, since $|x^{E}_i|$ and $|x^{D}_i|$ are not correlated, it is possible, at least in principle, to simultaneously generate viable masses for down-type quarks and charged leptons. The vector-like leptons and quarks from the $(\bm{5}_4,\overline{\bm{5}}_4)$ pair have masses $m_4^E=\sqrt{|M^{E}_1|^2+|M^{E}_2|^2+|M^{E}_3|^2+|M^{E}_4|^2}$ and $m_4^D=\sqrt{|M^{D}_1|^2+|M^{D}_2|^2+|M^{D}_3|^2+|M^{D}_4|^2}$, respectively. 

We clearly do not address the origin of neutrino mass at this stage. Our study, however, and all correlations we generate next will be valid for all $SU(5)$ models that do include neutrino mass generation but do not directly affect the charged fermion sector~\cite{Dorsner:2005fq,Bajc:2006ia}. We provide two realistic GUT models that allow for light vector-like states and incorporate neutrino masses in Section~\ref{MODELS}.
 
The parameters that are {\it a priori} unknown in our framework are $m_i$ and $|x^{E,D}_i|$, $i=1,2,3$. We thus have nine parameters to explain six experimentally measured masses, i.e., $m_i^E$ and $m_i^D$, $i=1,2,3$. At this stage the masses of vector-like leptons and quarks contribute only indirectly through the parameters $|x^{E}_i|$ and $|x^{D}_i|$ towards the masses of the matter fields. It turns out, however, that it is not trivial to simultaneously satisfy all six equations that relate $m_i^E$ and $m_i^D$ with $m_i$ and $|x^{E,D}_i|$, $i=1,2,3$. For example, there is no solution for $m_1 m_2 m_3 \leq m^E_1 m^E_2 m^E_3$. (The observed masses, when propagated to the GUT scale, yield $m^E_1 m^E_2 m^E_3 > m^D_1 m^D_2 m^D_3$. See, for example, Table~\ref{table:1}. The charged lepton masses hence play a more prominent role than the down-type quark masses in Eq.~\eqref{eq:3}.) In fact, there exists a number of additional constraints that originate from Eqs.~\eqref{eq:1},~\eqref{eq:2} and~\eqref{eq:3} on the parameters $m_i$, $i=1,2,3$. Here we elaborate on two of them. 

Since the observed masses of charged leptons and down-type quarks exhibit a rather strong hierarchy, we can safely neglect terms that are proportional to $m^{E,D}_1$ on the left-hand sides of Eqs.~\eqref{eq:1} and~\eqref{eq:2}. If we do that, we can rewrite Eq.~\eqref{eq:2} to read
\begin{equation}
\label{eq:2a}
(m^{E,D}_2 m^{E,D}_3)^2\approx \frac{m_2^2 m_3^2 (1+|x^{E,D}_1|^2)}{1+|x^{E,D}|^2}.
\end{equation}
Here we also neglect terms proportional to $m_1$ on the right-hand side of Eq.~\eqref{eq:2}. This then allows us to obtain an approximate equality $(m^{E,D}_1)^2 \approx m_1^2/(1+|x^{E,D}_1|^2)$, once we use Eq.~\eqref{eq:3}. Since $m^{D}_1>m^{E}_1$ we finally get   
\begin{equation}
m_1^D \lesssim m_1.
\end{equation}
This result implies that it is impossible to simultaneously solve all six equations if $m_1$ is below the mass of the down quark. 

Let us now neglect terms that are proportional to $m^{E,D}_1$ and $m^{E,D}_2$ in Eq.~\eqref{eq:1}. We accordingly neglect $m_1$ and $m_2$ in Eq.~\eqref{eq:1} to find that 
\begin{equation}
(m^{E,D}_3)^2\approx \frac{m_3^2 (1+|x^{E,D}_1|^2+|x^{E,D}_2|^2)}{1+|x^{E,D}|^2}.
\end{equation}
This result, when combined with Eq.~\eqref{eq:2a}, yields 
\begin{equation}
m_2^2 \approx (m^{E,D}_2)^2 \left(1+\frac{|x^{E,D}_2|^2}{1+|x^{E,D}_1|^2}\right).
\end{equation}
This approximate equality implies that it is impossible to simultaneously solve all six equations if $m_2$ is below the muon mass, i.e., $m_2^E \lesssim m_2$, since $m_2^E>m_2^D$.

Our numerical procedure confirms these two relations. We actually find the following set of inequalities to be satisfied: $m_1^D \lesssim m_1 \lesssim m_2^D$, $m_2^E \lesssim m_2 \leq m_3^D$ and $m_3^E \lesssim m_3$. Note that perturbativity considerations place an upper bound on $m_3$. However, the flavor physics constraints keep all Yukawa couplings well below that threshold as we show later on. Note also that Eqs.~\eqref{eq:1},~\eqref{eq:2} and~\eqref{eq:3} are invariant under the exchange $(m_i, |x^{E,D}_i|) \leftrightarrow (m_j, |x^{E,D}_j|)$, $i,j=1,2,3$. To account for that we consider only the following ordering: $m_1<m_2<m_3$.

What we want to study is how much of the allowed parameter space spanned by $m_i$ and $|x^{E,D}_i|$ satisfies applicable low-energy constraints, once we ask for the vector-like states to be light. To accomplish that goal we implement the following numerical procedure. We first specify $m_i^{E,D}$, $i=1,2,3$, at the GUT scale to be our input. Relevant values of $m_i^{E,D}$ we use are summarized in Table~\ref{table:1}. We consider a scenario without supersymmetry and take the GUT scale to be $m_\mathrm{GUT}=10^{16}$\,GeV. The running of masses is performed under the assumptions specified in Ref.~\cite{Dorsner:2011ai}. The $m_\mathrm{GUT}$ values shown in Table~\ref{table:1} are to be understood as representative values that would change if one changes the GUT scale and/or introduces additional particles. Only central values for masses of down-type quarks and charged leptons are considered in our study. 

Once the input is defined, we vary the parameters $m_i$ until we numerically obtain particular values of $|x_i^{E}|$'s that simultaneously satisfy Eqs.~\eqref{eq:1},~\eqref{eq:2} and~\eqref{eq:3}. We then fix the $m_i$'s to these values and proceed to find viable solutions that relate $m_i^{D}$ and $|x_i^{D}|$. What we end up with are viable sets of $m_i$ values and associated values of the parameters $|x_i^{E,D}|$ that yield realistic masses for down-type quarks and charged leptons at the GUT scale. In other words, we find all possible values of $m_i$ and $|x_i^{E,D}|$ that satisfy Eqs.~\eqref{eq:1},~\eqref{eq:2} and~\eqref{eq:3} for a given set of $m_i^{E,D}$, $i=1,2,3$. Finally, we test whether these solutions also satisfy low-energy constrains after we specify the masses of the vector-like states. $m_4^{E,D}$ need to be specified in order for us to determine unitary transformations that bring the $4 \times 4$ matrices $M_E$ and $M_D$, explicitly shown in Eq.~\eqref{4times4}, into a diagonal form that corresponds to the fermion mass eigenstate basis. It is these unitary transformations that enter low-energy considerations as we demonstrate in section~\ref{CONSTRAINTS}. Once we specify $m_4^{D}$ ($m_4^{E}$), we numerically determine all entries of the matrix $X^d$ ($Y^e$)  and test these entries against the constraints presented in Table~\ref{Zcoups} (Table~\ref{lepZ}). For example, to find $Y^e$ we first construct a real normal matrix $(M^T_E M_E)$ that we diagonalize with a congruent transformation $E_R (M^T_E M_E) E^T_R=(M^T_E M_E)^{\mathrm{diag}}$. This then allows us to define $|Y^e_{ij}|=|\sum_{k=1}^{3}(E_R)_{ik} (E_R)_{jk}-\delta_{ij}|/2$.

Note that the unitary transformations we find numerically that act on the matrices $M_E$ and $M_D$ are valid at the GUT scale. On the other hand, the constraints we want to impose on the entries of $X^d$ and $Y^e$ are valid at the low-energy scale. We, however, opt not to run numerically obtained entries of $X^d$ and $Y^e$ to low energies, since the angles that enter unitary transformations that define them are required to be small to satisfy low-energy constraints and should thus not change substantially through the running. 

The vector-like states we consider can be either quarks or leptons. We accordingly study and present both cases separately. In particular, when we consider the scenarios with light quark and lepton vector-like states we take $m_4^{D}=800$\,GeV and $m_4^{E}=400$\,GeV, respectively. These masses are allowed by direct searches for the vector-like states. Again, to numerically determine $X^d$ ($Y^e$) we need to specify $m_4^{D}$ ($m_4^{E}$).

We choose to present the outcome of our numerical analysis in form of plots of $m_2$ vs.\ $m_3$ for a given value of $m_1$. In other words, we have regions of constant $m_1$ in the $m_2$--$m_3$ plane that represent a phenomenologically viable parameter space. Every point within that region is associated with a unique set of values of $|x_i^{E,D}|$'s that were generated for a given set of $m_i$, $i=1,2,3$, that satisfy Eqs.~\eqref{eq:1},~\eqref{eq:2} and~\eqref{eq:3}.
\begin{table}[h]
\begin{center}
\begin{tabular}{|l|l|l|l|l|l|l|}
\hline
$\mu$ & $m_1^D(\mu)$ (GeV) & $m_2^D(\mu)$ (GeV) & $m_3^D(\mu)$ (GeV) & $m_1^E(\mu)$ (GeV) & $m_2^E(\mu)$ (GeV) & $m_3^E(\mu)$ (GeV)\\
\hline
\hline
$m_Z$ & 0.00350 & 0.0620
& 2.890 & 0.000487 & 0.103 & 1.75\\
$m_\mathrm{GUT}$ & 0.00105 & 0.0187
& 0.782 & 0.000435 & 0.092 & 1.56\\
\hline
\end{tabular}
\end{center}
\caption{Central values for masses of the SM down-type quarks $m_i^D(\mu)$ and charged leptons $m_i^E(\mu)$, $i=1,2,3$, at $\mu=m_Z$ and $\mu=m_\mathrm{GUT}$, where $m_\mathrm{GUT}=10^{16}$\,GeV.} \label{table:1}
\end{table}

\begin{figure}[htb]
\begin{center}
\includegraphics[width=11.5cm]{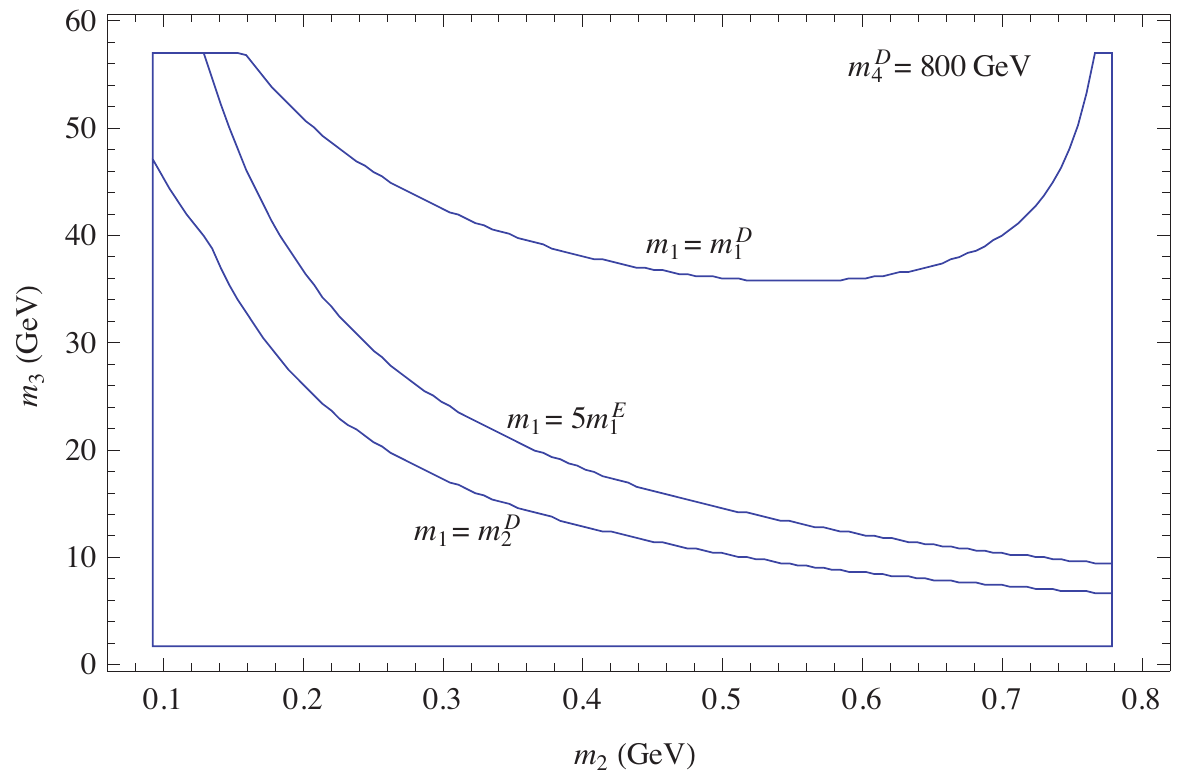}
\end{center}
\caption{Allowed parameter space in the $m_2$--$m_3$ plane for $m_4^{D}=800$\,GeV as a function of $m_1$. The contours define viable regions generated for $m_1=m^D_2$, $m_1=5 m^E_1$ and $m_1=m^D_1$, starting with the innermost one. Regions outside of the contours are excluded.}
\label{fig:q_1}
\end{figure}

\begin{figure}[htb]
\begin{center}
\includegraphics[width=11.5cm]{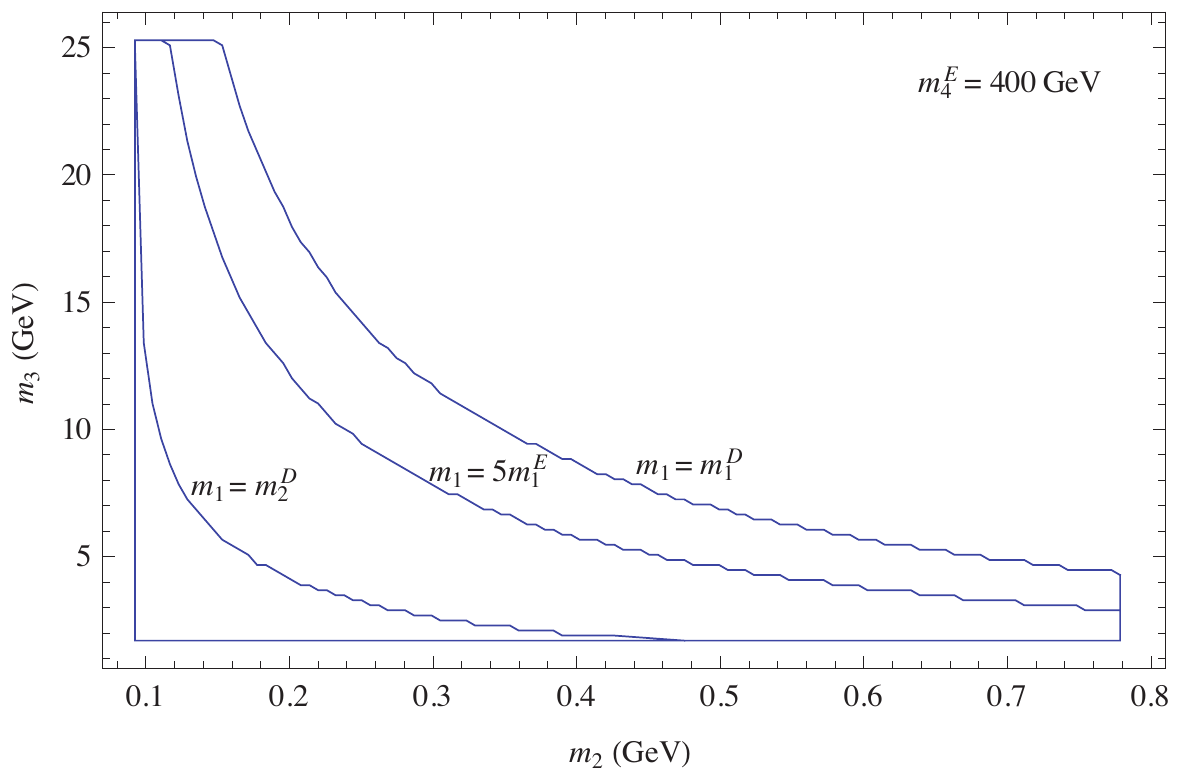}
\end{center}
\caption{Allowed parameter space in the $m_2$--$m_3$ plane for $m_4^{E}=400$\,GeV. The contours define viable regions generated for $m_1=m^D_2$, $m_1=5 m^E_1$ and $m_1=m^D_1$, starting with the innermost one. Regions outside of the contours are excluded.}
\label{fig:l_1}
\end{figure}

We show in Figs.~\ref{fig:q_1} and~\ref{fig:l_1} the parameter space allowed by low-energy constraints in the $m_2$--$m_3$ plane for $m_4^{D}=800$\,GeV and $m_4^{E}=400$\,GeV, respectively. The contours bound viable regions of constant $m_1$ in the $m_2$--$m_3$ plane that yield satisfactory fermion masses. We opt to present regions with $m_1=m^D_2$, $m_1=5 m^E_1$ and $m_1=m^D_1$. Recall, there exists no solution for $m_1<m^D_1$ or $m_1>m^D_2$. 

One can see from Figs.~\ref{fig:q_1} and~\ref{fig:l_1} that the allowed region for a given value of $m_1$ in the case of a light vector-like quark state is significantly less constrained with regard to the case when the light state is a vector-like lepton. This was to be expected, since the experimental constraints from observables that concern flavor physics effects in the charged lepton sector are much more stringent than in the down-type quark sector. The allowed region for the $m_4^D=800$\,GeV case, for fixed $m_1$, becomes comparable in size to the region that corresponds to the case of a light vector-like lepton state only when $m_4^E\approx1.2$\,TeV. It is interesting to note that $m_i$, $i=1,2,3$, exist only within narrow ranges once the low-energy constraints are implemented. For example, once the vector-like leptons are taken to be light,  $m_1$ and $m_3$ can change, at most, by about a factor of twenty whereas $m_2$ can change by about a factor of eight. (See Fig.~\ref{fig:l_1}.) It is also clear from our numerical study that $m_2^E \lesssim m_2 \leq m_3^D$ and $m_3^E \lesssim m_3$.

The shape of the available parameter space shown in Figs.~\ref{fig:q_1} and~\ref{fig:l_1} has a very simple interpretation. The size of the bounded area shrinks as $m_1$ grows simply due to the fact that any departure of $m_1$ from its ``natural'' value requires more substantial rotations of the $4 \times 4$ matrices $M_D$ and $M_E$ to correctly account for the masses for the first generation of down-type quarks and charged leptons, respectively. These rotations, on the other hand, need to be small if one is to satisfy existing low-energy constraints, especially when the light vector-like fermions are leptons. The same effect is evident with regard to departures of $m_2$ and $m_3$ from their preferred values that are set by the mass scales of the second and third generation of down-type quarks and charged leptons, respectively. Clearly, the most natural and hence the least constrained part of the available parameter space is the one where $m_i \sim m_i^{E,D}$, $i=1,2,3$.


\section{Predictions}
\label{PREDICTIONS}

All viable extensions that represent minimal departures from the original $SU(5)$ setup are very predictive~\cite{Dorsner:2005fq,Bajc:2006ia,Dorsner:2012uz} with regard to proton decay. The same is true for the framework under consideration that includes only one extra vector-like pair of fields with regard to the Georgi-Glashow model~\cite{Georgi:1974sy} as we show next.

The scenario with an extra vector-like pair $(\bm{5}_4,\overline{\bm{5}}_4)$ yields all unitary transformations that are necessary to go from an arbitrary flavor basis to the mass eigenstate one in terms of three parameters. These parameters, i.e., $m_1$, $m_2$ and $m_3$, suffice to describe all viable redefinitions of the SM down-type quark and the SM charged lepton fields as we demonstrated in Section~\ref{FRAMEWORK}. The description of all unitary transformations in the up-type quark sector, on the other hand, requires no additional parameters for the following reason. The $SU(5)$ invariant operator that generates all up-type quark masses guarantees the symmetric nature of the relevant mass matrix. This fact allows one to relate rotations in the left-handed sector of up-type quarks to rotations in the right-handed sector of up-type quarks. Moreover, it is the CKM mixing matrix that provides the link between left-handed rotations in the up- and the down-type quark sectors. To establish this connection we use the following CKM parameters: $\lambda = 0.22535$, $A = 0.811$, $\bar{\rho} = 0.131$ and $\bar{\eta} = 0.345$~\cite{Beringer:1900zz}.

The framework thus yields accurate proton decay signatures for two-body decays of the proton through gauge boson exchange in terms of only three parameters that are already constrained to be within very narrow ranges. Note that the exact mechanism of the neutrino mass generation cannot affect these predictions since the relevant decay amplitudes do not refer to the neutrino mixing parameters. In any case, the fact that the transformations in the charged lepton sector are known in terms of $m_i$, $i=1,2,3$, allows one to reconstruct unitary transformations in the neutrino sector via the Pontecorvo-Maki-Nakagawa-Sakata (PMNS) mixing matrix.

The gauge coupling strength at the GUT scale is $\alpha_\mathrm{GUT}=g^2_\mathrm{GUT}/(4 \pi)$ and the mass of the gauge boson that mediates proton decay corresponds to $m_\mathrm{GUT}$ in $SU(5)$. We take $\alpha_\mathrm{GUT}=0.033$~\cite{Dorsner:2011ai} and $m_\mathrm{GUT}=4 \times 10^{15}$\,GeV for definiteness in our numerical analysis of partial proton decay widths. Recall, $m_\mathrm{GUT}$ can be identified with the mass of proton decay mediating gauge bosons in $SU(5)$. The partial lifetimes we present scale with $\alpha_\mathrm{GUT}^{-1}$ ($m_\mathrm{GUT}$) to the second (fourth) power and can thus be easily recalculated for different values of these two parameters. To generate proton decay predictions we furthermore use $\hat \alpha=-0.0112$\,GeV$^3$~\cite{Aoki:2008ku}, where $\hat \alpha$ is the relevant nucleon matrix element. The leading-log renormalization corrections of the $d=6$ operator coefficients are taken to be $A_{S\,L} = 2.6$ and $A_{S\,R} = 2.4$~\cite{Dorsner:2009mq} and the exact dependence of the decay amplitudes on unitary transformations is taken from Ref.~\cite{FileviezPerez:2004hn}.

\begin{figure}[htb]
\begin{center}
\includegraphics[width=10.5cm]{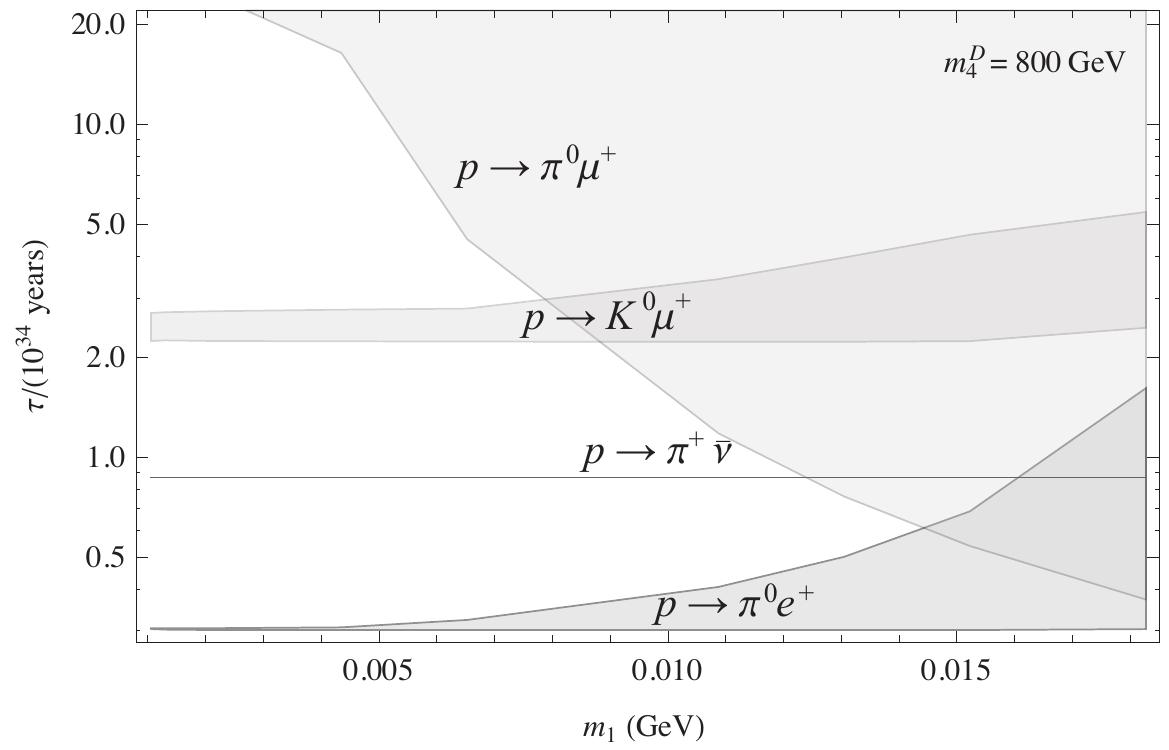}
\end{center}
\caption{Plot of $m_1$ vs.\ ${\tau}_{p \rightarrow \pi^0 e^+}$, ${\tau}_{p \rightarrow \pi^0 \mu^+}$, ${\tau}_{p \rightarrow \pi^+ \bar{\nu}}$ and ${\tau}_{p \rightarrow K^0 \mu^+}$. The bands represent the ranges of predicted values for partial proton decay lifetimes for associated decay channels for the $m_4^D=800$\,GeV case.}
\label{fig:p_1_D}
\end{figure}

\begin{figure}[htb]
\begin{center}
\includegraphics[width=10.5cm]{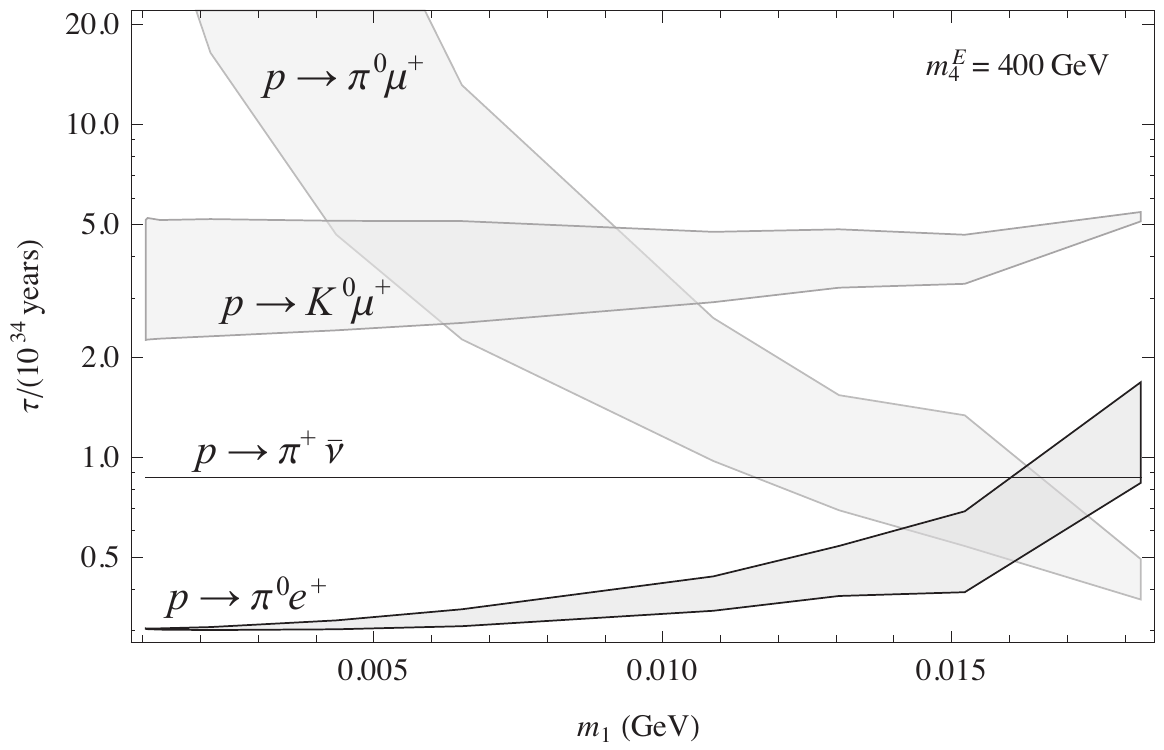}
\end{center}
\caption{Plot of $m_1$ vs.\ ${\tau}_{p \rightarrow \pi^0 e^+}$, ${\tau}_{p \rightarrow \pi^0 \mu^+}$, ${\tau}_{p \rightarrow \pi^+ \bar{\nu}}$ and ${\tau}_{p \rightarrow K^0 \mu^+}$. The bands represent the ranges of predicted values for partial proton decay lifetimes for associated decay channels for the $m_4^E=400$\,GeV case.}
\label{fig:p_1_E}
\end{figure}

We present results for the four proton decay channels that turn out to be the most relevant ones in Figs.~\ref{fig:p_1_D} and~\ref{fig:p_1_E}. These channels are $p \rightarrow \pi^0 e^+$, $p \rightarrow \pi^0 \mu^+$, $p \rightarrow \pi^+ \bar{\nu}$ and $p \rightarrow K^0 \mu^+$. The current experimental limits are ${\tau}_{p \rightarrow \pi^0 e^+}> 1.3 \times 10^{34}$\,years~\cite{Nishino:2012ipa}, ${\tau}_{p \rightarrow \pi^0 \mu^+}>  1.1 \times 10^{34}$\,years~\cite{Nishino:2012ipa}, ${\tau}_{p \rightarrow \pi^+ \bar{\nu}}>  3.9 \times 10^{32}$\,years~\cite{Abe:2013lua} and ${\tau}_{p \rightarrow K^0 \mu^+}>  1.6 \times 10^{33}$\,years~\cite{Regis:2012sn}. (All other two-body decay modes of the proton can be safely neglected for all practical purposes.) The parameter points we use to generate proton decay predictions in Figs.~\ref{fig:p_1_D} and~\ref{fig:p_1_E} correspond to the $m_4^D=800$\,GeV and $m_4^E=400$\,GeV cases shown in Figs.~\ref{fig:q_1} and~\ref{fig:l_1}, respectively. We plot minimal and maximal values of predicted partial lifetimes for four proton decay channels in Figs.~\ref{fig:p_1_D} and~\ref{fig:p_1_E} for all allowed values of parameter $m_1$: $m_1 \in ( m_1^D, m_2^D)$. Note that we extract relevant unitary transformations using reduced forms of the original matrices $M_E$ and $M_D$ that are given in Eq.~\eqref{4times4}. Namely, we bring the effective $3 \times 3$ matrices $\hat{M}_E$ and $\hat{M}_D$ of the SM charged leptons and down-type quarks to diagonal form. $\hat{M}_E$ explicitly reads~\cite{Babu:2012pb} 
\begin{equation}
\label{3times3}
\hat{M}_E = \left( \begin{array}{ccc}
\frac{m_1}{\sqrt{1+|x_1^E|^2}} & 0   & 0 \\
-\frac{m_1 |x_1^E| |x_2^E|}{\sqrt{1+|x_1^E|^2}\sqrt{1+|x_1^E|^2+|x_2^E|^2}} & m_2 \sqrt{\frac{1+|x_1^E|^2}{1+|x_1^E|^2+|x_2^E|^2}} & 0 \\
-\frac{m_1 |x_1^E| |x_3^E|}{\sqrt{1+|x_1^E|^2+|x_2^E|^2}\sqrt{1+|x^E|^2}} & -\frac{m_2 |x_2^E| |x_3^E|}{\sqrt{1+|x_1^E|^2+|x_2^E|^2}\sqrt{1+|x^E|^2}} & m_3 \sqrt{\frac{1+|x_1^E|^2+|x_2^E|^2}{1+|x^E|^2}} 
\end{array} \right).
\end{equation}
To obtain $\hat{M}_D$ from $\hat{M}_E$ all one needs to do is to replace $|x_i^E|$ with $|x_i^D|$, $i=1,2,3$, in Eq.~\eqref{3times3} and transpose the resulting mass matrix. 

The bands in Figs.~\ref{fig:p_1_D} and~\ref{fig:p_1_E} represent predicted values for associated proton decay channels when $m_4^D=800$\,GeV and $m_4^E=400$\,GeV, respectively. They reflect the dependence of the decay amplitudes on unitary transformations of the quark and lepton fields that, in turn, depend on $m_i$, $i=1,2,3$ parameters. The most important observation is the possibility that one can pinpoint the value of $m_1$ for a fixed value of either $m_4^D$ or $m_4^E$ through the proton decay signatures in this framework. 

For definiteness, let us address the $m_4^E=400$\,GeV case. For small values of $m_1$, it is the $p \rightarrow \pi^0 e^+$ signature that dominates. It is followed by $p \rightarrow \pi^+ \bar{\nu}$, $p \rightarrow K^0 \mu^+$ and $p \rightarrow \pi^0 \mu^+$. For larger values of $m_1$, on the other hand, the $p \rightarrow \pi^0 \mu^+$ signature starts to dominate over $p \rightarrow \pi^0 e^+$, with the processes $p \rightarrow \pi^0 e^+$ and $p \rightarrow \pi^+ \bar{\nu}$ being of the same strength. Another nice feature of the framework is the constancy of the $p \rightarrow \pi^+ \bar{\nu}$ decay amplitude. This is expected since the mass matrix of the up-type quark sector is symmetric and the sum over all neutrino flavors removes all dependence on the unitary transformations~\cite{FileviezPerez:2004hn}. Finally, we see that the amplitudes for both the $p \rightarrow \pi^0 e^+$ and the $p \rightarrow K^0 \mu^+$ modes vary only slightly with regard to the parameters of the proposed extension. Clearly, the low-energy phenomenology allows for the presence of only those transformations that correspond to small changes in the angles of rotations that enter redefinitions of the quark and lepton fields. This, again, is reflected in the narrow widths of the allowed bands for predicted proton decay signatures. 

\begin{figure}[htb]
\begin{center}
\includegraphics[width=10.5cm]{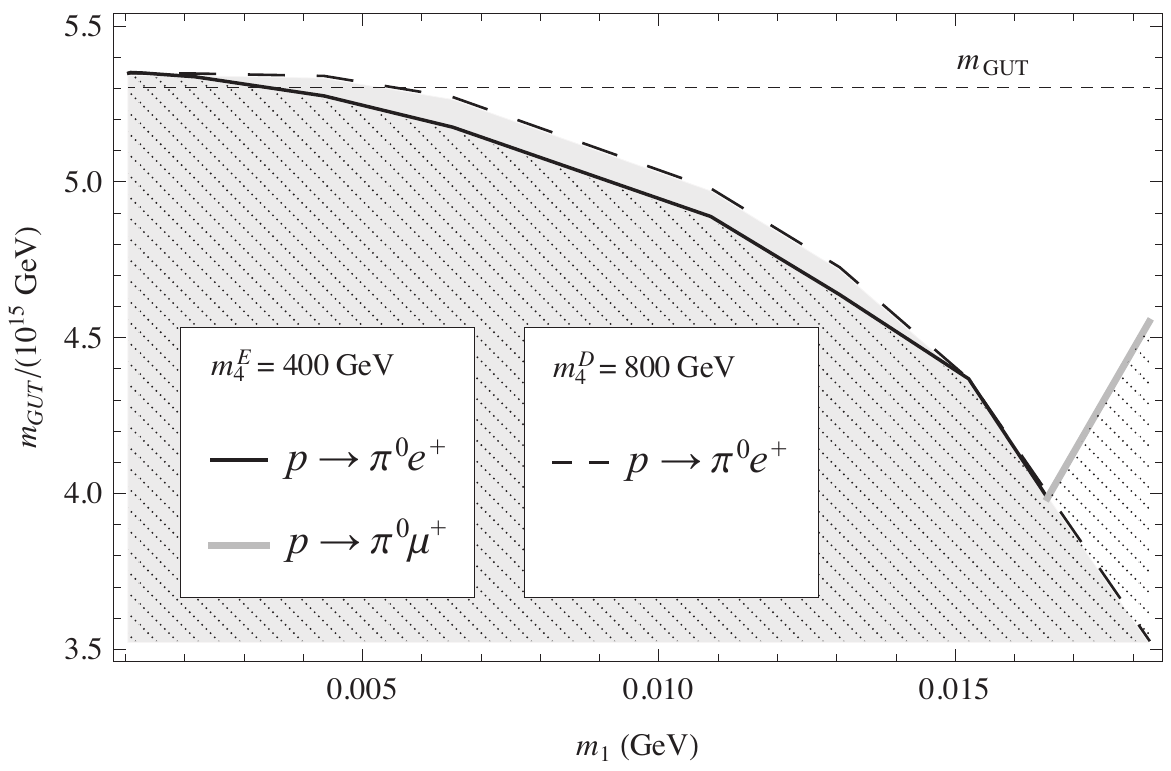}
\end{center}
\caption{Plot of $m_1$ vs.\ $m_\mathrm{GUT}$. The excluded (dotted) region is bounded by predictions for ${\tau}_{p \rightarrow \pi^0 e^+}$ (black curve) and ${\tau}_{p \rightarrow \pi^0 \mu^+}$ (grey curve) for the $m_4^E=400$\,GeV case. The excluded (shaded) region for the $m_4^D=800$\,GeV case is bounded by ${\tau}_{p \rightarrow \pi^0 e^+}$ (dashed curve).}
\label{fig:p_2}
\end{figure}

We can use the predictions for the partial decay widths we displayed in Figs.~\ref{fig:p_1_D} and~\ref{fig:p_1_E} to find conservative lower bounds on the GUT scale in this framework for the $m_4^D=800$\,GeV and $m_4^E=400$\,GeV cases, respectively. We present these bounds in Fig.~\ref{fig:p_2}.
Note that the bound on $m_\mathrm{GUT}$ is set by two distinct proton decay channels for the $m_4^E=400$\,GeV case. Namely, for small (large) values of $m_1$ the bound is set by $p \rightarrow \pi^0 e^+$ ($p \rightarrow \pi^0 \mu^+$). This bound is shown in Fig.~\ref{fig:p_2} as a solid curve. In the $m_4^D=800$\,GeV case the bound is set solely by the $p \rightarrow \pi^0 e^+$ channel. This bound is shown in Fig.~\ref{fig:p_2} as a dashed curve. The extraction of these bounds is done for $\alpha_\mathrm{GUT}^{-1}=35.0$ for reasons that are explained in the next section. The origin of a horizontal dashed line in Fig.~\ref{fig:p_2} will also be addressed there. 


\section{MODELS}
\label{MODELS}

The $SU(5)$ framework we study in Sections~\ref{FRAMEWORK} and~\ref{PREDICTIONS} contains only one vector-like pair $(\bm{5}_4,\overline{\bm{5}}_4)$ of matter fields in addition to the particle content of the Georgi-Glashow $SU(5)$ model. It is easy to show that this scenario cannot possibly provide a phenomenologically viable unification of gauge couplings. Namely, the highest possible value of the GUT scale $m_\mathrm{GUT}$ cannot exceed $10^{14.18}$\,GeV if we assume that $m_4^E=400$\,GeV. This value of the GUT scale is already excluded by current experimental data on stability of matter. (See, for example, Fig.~\ref{fig:p_2}.) 

The lowest mass one can have for the vector-like down-type quarks, on the other hand, turns out to be $m_4^D=10^{8.50}$\,GeV with the corresponding GUT scale of $m_\mathrm{GUT}=10^{13.96}$\,GeV. This implies that we cannot even consider having the vector-like down-type quarks in this framework at the electroweak scale if we require unification regardless of the proton decay constraints. (These numerical results are based on the one-loop gauge coupling unification analysis and are obtained with the following input parameters: $\sin^2 \theta_W (M_Z) = 0.23122$, $\alpha(M_Z) = 1/127.906$ and $\alpha_3(M_Z) = 0.1184$. We consider only central values of these parameters and allow the masses of relevant multiplets to vary freely in the allowed parameter space.)

One might thus object that our scenario is not self-consistent and that our predictions for proton decay modes are not relevant. We stress, however, that the predictions we generate in Section~\ref{PREDICTIONS} are valid in any $SU(5)$ scenario where the mismatch between the down-type quarks and the charged leptons is addressed solely through the use of a vector-like pair $(\bm{5}_4,\overline{\bm{5}}_4)$ of matter fields.

What would then be a realistic setting for our scenario? Firstly, the model should not rely on higher-dimensional operators to accommodate the observed fermion masses. This would make the presence of the vector-like pair superfluous. Secondly, a viable setting should not include a $45$-dimensional scalar representation for the same reason. Namely, the introduction of an additional $SU(2)$ doublet in a $45$-dimensional scalar representation beside the one in the $5$-dimensional scalar representation would allow one to address the mismatch between the down-type quarks and the charged leptons~\cite{Georgi:1979df} without a need to resort to vector-like fermion fields. Thirdly, one should not introduce additional vector-like fields that would mix with either the down-type quarks or the charged leptons. Next, we present two simple extensions of the minimal Georgi-Glashow $SU(5)$ with vector-like fermions that automatically satisfy all three aforementioned requirements and allow for gauge coupling unification.


\subsection{Scenario I}
The first extension we propose uses one $50$-dimensional scalar representation that can thus couple only to the matter fields through the $\bm{10}_i \bm{10}_j \overline{\bm{50}}$ term, where $i,j=1,2,3$. This means that the vector-like fermion couplings to matter are not disturbed in any way by the introduction of this scalar representation. The proposed addition also does not affect the symmetric nature of the up-type quark mass matrix. 

We denote the components of the $50$-dimensional representation with $\bm{50}^{\alpha \beta \gamma \delta}$, $\alpha, \beta, \gamma, \delta=1,\ldots,5$, where the SM gauge group decomposition reads $\bm{50}= (\bm{8},\bm{2},1/2)\oplus
(\bm{6},\bm{1}, 4/3) \oplus (\overline{\bm{6}},\bm{3},-1/3)
\oplus (\overline{\bm{3}}, \bm{2}, -7/6) \oplus (\bm{3}, \bm{1}, -1/3) \oplus (\bm{1},
\bm{1}, -2)=\Phi_1 \oplus \Phi_2 \oplus \Phi_3 \oplus \Phi_4 \oplus \Phi_5 \oplus \Phi_6$. The relevant properties of this representation are $\bm{50}^{\alpha \beta \gamma \delta} = -\bm{50}^{\beta \alpha \gamma \delta} = -\bm{50}^{\alpha \beta  \delta \gamma}=\bm{50}^{\gamma \delta \alpha \beta}$ and $\epsilon_{\zeta \alpha \beta \gamma \delta} \bm{50}^{\alpha \beta \gamma \delta}=0$, where $\zeta=1,\ldots,5$. Recall, other scalar representations already present in our scenario are $\bm{5}= (\bm{1},\bm{2},1/2)\oplus(\bm{3},\bm{1},-1/3) = \Psi_D \oplus \Psi_T$ and 
$\bm{24} = (\bm{8},\bm{1},0) \oplus (\bm{1},\bm{3},0) \oplus (\bm{3},\bm{2},-5/6) \oplus
(\overline{\bm{3}},\bm{2},5/6) \oplus (\bm{1},\bm{1},0)= \Sigma_8 \oplus \Sigma_3 \oplus \Sigma_{(3,2)} \oplus \Sigma_{(\bar{3}, 2)} \oplus \Sigma_{24}$. We accordingly find four independent $SU(5)$ invariant contractions of the $50$-dimensional representation that generate the masses of its components. (Note that we completely neglect all contributions from the electroweak VEV towards the masses of the SM gauge group multiplets in the $50$-dimensional representation.) These contractions are $m^2 \bm{50}^{\alpha \beta \gamma \delta} \overline{\bm{5}}_{\alpha \beta \gamma \delta}$, $m' \bm{50}^{\alpha \beta \gamma \delta} \bm{24}^\zeta_\delta \overline{\bm{50}}_{\zeta \beta \gamma \alpha}$, $\lambda \bm{50}^{\alpha \beta \gamma \delta} \bm{24}^\zeta_\delta \bm{24}^\epsilon_\zeta \overline{\bm{50}}_{\epsilon \gamma \alpha \beta}$ and $\lambda' \bm{50}^{\alpha \beta \gamma \delta} \bm{24}^\zeta_\alpha \bm{24}^\epsilon_\beta \overline{\bm{50}}_{\gamma \delta \zeta \epsilon}$, where $\alpha, \beta, \gamma, \delta, \zeta, \epsilon=1,\ldots,5$ represent $SU(5)$ indices. We accordingly find
\begin{align}
m^2_{\Phi_1}&=m^2 + \frac{3}{8} m' \sigma - \frac{21}{4} \lambda \sigma^2 - \lambda' \sigma^2,\\
m^2_{\Phi_2}&=m^2 + m' \sigma - 4 \lambda \sigma^2 + 4 \lambda' \sigma^2,\\
m^2_{\Phi_3}&=m^2 - \frac{1}{4} m' \sigma - \frac{13}{2} \lambda \sigma^2 - 6 \lambda' \sigma^2,\\
m^2_{\Phi_4}&=m^2 - \frac{7}{8} m' \sigma - \frac{31}{4} \lambda \sigma^2 + \frac{3}{2} \lambda' \sigma^2,\\
m^2_{\Phi_5}&=m^2 - \frac{1}{4} m' \sigma - \frac{13}{2} \lambda \sigma^2 + \frac{7}{3} \lambda' \sigma^2,\\
m^2_{\Phi_6}&=m^2 - \frac{3}{2} m' \sigma - 9 \lambda \sigma^2 + 9 \lambda' \sigma^2,
\end{align}
where $\sigma$ is the VEV of the adjoint representation of $SU(5)$ defined in Section~\ref{FRAMEWORK}.

Clearly, the $m^2$ term represents a common contribution towards $m^2_{\Phi_i}$, $i=1,\ldots,6$. This leaves us with three independent parameters that control the mass splittings between the SM gauge group components of the $50$-dimensional representation. These mass splittings are important if we want to provide a viable gauge coupling unification with light vector-like fermions. If one eliminates these independent parameters in favor of $m_{\Phi_4}$, $m_{\Phi_5}$ and $m_{\Phi_6}$ one obtains the following mass relations 
\begin{align}
\label{eq:Phi1}
m^2_{\Phi_1}&=\frac{1}{2}(3 m^2_{\Phi_5}-m^2_{\Phi_6}),\\\label{eq:Phi2}
m^2_{\Phi_2}&=(3 m^2_{\Phi_5}-2 m^2_{\Phi_4}),\\\label{eq:Phi3}
m^2_{\Phi_3}&=(2 m^2_{\Phi_4}-m^2_{\Phi_6}),
\end{align}
that should hold at the GUT scale.

The rest of the scalar sector is such that it allows all possible mass splittings between various multiplets in $\bm{24}$ and $\bm{5}$. An additional constraint comes from experimental results on proton decay that set a lower limit on the masses of both $\Psi_T$ and $\Phi_5$. We take this limit to be $m_{\Psi_T}, m_{\Phi_5} > 3 \times 10^{11}$\,GeV~\cite{Dorsner:2012uz}. Also, we require that $m_{\Sigma_8}>10^5$\,GeV~\cite{Bajc:2006ia} and that $4 \times 10^2\,\mathrm{GeV} \leq m_{\Phi_1},m_{\Phi_2},m_{\Phi_3},m_{\Phi_4},m_{\Phi_6},m_{\Sigma_3}  \leq m_\mathrm{GUT}$. Note that $\Sigma_{(3,2)}$ and $\Sigma_{(\bar{3}, 2)}$ are eaten by proton decay mediating gauge bosons and that their masses are thus identified with $m_\mathrm{GUT}$. We also bound $m_{\Phi_5}$ from above by $m_\mathrm{GUT}$. This ensures that the gauge couplings stay unified above the GUT scale.

We consider two special cases with regard to unification to demonstrate the viability of the proposed extension. We first maximize $m_\mathrm{GUT}$ under the assumption that $m_4^E=400$\,GeV and $m_4^D \geq 800$\,GeV requiring that the exact gauge coupling unification takes place at the one-loop level. We vary the masses of $\Sigma_8$, $\Sigma_3$, $\Psi_T$, $\Phi_1$, $\Phi_2$, $\Phi_3$, $\Phi_4$, $\Phi_5$ and $\Phi_6$ within the allowed ranges taking into account Eqs.~\eqref{eq:Phi1},~\eqref{eq:Phi2} and~\eqref{eq:Phi3}.
This procedure yields $m_\mathrm{GUT}=10^{16.25}$\,GeV. Again, this is the highest possible value of the GUT scale under the assumption that $m_4^E=400$\,GeV. This immediately tells us that the proposed extension is viable since the GUT scale exceeds the lower limits we present in Fig.~\ref{fig:p_2}. The maximum for $m_\mathrm{GUT}$ is achieved for the following values of masses: $m_{\Sigma_3}=10^{16.25}$\,GeV, $m_{\Sigma_8}=10^{5.00}$\,GeV,  $m_{\Psi_T}=3 \times 10^{11}$\,GeV,  $m_{\Phi_1}=10^{14.44}$\,GeV,  $m_{\Phi_2}=10^{14.59}$\,GeV,  $m_{\Phi_3}=10^{8.58}$\,GeV,  $m_{\Phi_4}=10^{15.51}$\,GeV,  $m_{\Phi_5}=10^{15.42}$\,GeV,  $m_{\Phi_6}=10^{15.66}$\,GeV, $m_4^D=800$\,GeV and $m_4^E=400$\,GeV. The corresponding value of the inverse of the gauge coupling at the GUT scale is $\alpha_\mathrm{GUT}^{-1}=27.2$. Another nice feature of this case is that the demand that $m_4^E$ is light also makes $m_4^D$ light, if one is to maximize the GUT scale. Again, any departure from the mass spectrum we present would decrease $m_\mathrm{GUT}$.

The second case is when we fix $m_4^D=800$\,GeV and demand that $m_4^E>1.2$\,TeV to make sure that the constraints we discuss in Section~\ref{FRAMEWORK} originate from the presence of the vector-like down-type quarks. The maximum value of the GUT scale is $m_\mathrm{GUT}=10^{16.83}$\,GeV and it is achieved for the following values of masses: $m_{\Sigma_3}=400$\,GeV, $m_{\Sigma_8}=10^{5}$\,GeV,  $m_{\Psi_T}= 10^{14.14}$\,GeV,  $m_{\Phi_1}=10^{16.68}$\,GeV,  $m_{\Phi_2}=10^{16.83}$\,GeV,  $m_{\Phi_3}=10^{8.59}$\,GeV,  $m_{\Phi_4}=10^{15.51}$\,GeV,  $m_{\Phi_5}=10^{15.60}$\,GeV,  $m_{\Phi_6}=10^{15.66}$\,GeV, $m_4^D=800$\,GeV and $m_4^E=10^{10.54}$\,GeV. We find that $\alpha_\mathrm{GUT}^{-1}=29.6$ in this particular case.

It is clear that both cases are viable and yield phenomenologically acceptable upper limits on the GUT scale. The most prominent feature of this scenario is the fact that $\Phi_3$ should be relatively light to maximize the GUT scale. This is easy to understand considering the fact that $(b_1-b_2)=-54/15$ and $(b_2-b_3)=+9/6$ for this field, where $b_i$, $i=1,2,3$, represent the one-loop $\beta$-function coefficients~\cite{Cheng:1973nv}. The field $\Phi_3$ thus efficiently unifies gauge coupling constants and, at the same time, increases the GUT scale. 

The model could furthermore accommodate experimental data on neutrino masses through the addition of at least two fermions that are singlets with regard to the SM gauge groups. This would not affect the unification considerations in any way. One could also add a $75$-dimensional scalar representation to implement the missing partner mechanism in $SU(5)$. This would only relax the constraints imposed by gauge coupling unification and proton decay that we have just considered. All in all, this is a phenomenologically viable extension of the minimal $SU(5)$ that allows for light vector-like fermions and does not affect the predictions presented in Section~\ref{PREDICTIONS}. 


\subsection{Scenario II}
The second extension we want to pursue is based on the addition of extra fermions to the minimal $SU(5)$ with a vector-like pair $(\bm{5}_4,\overline{\bm{5}}_4)$ of matter fields. Namely, we opt to add two $24$-dimensional representations of fermions --- $\bm{24}_i=(\bm{8},\bm{1},0)_i \oplus (\bm{1},\bm{3},0)_i \oplus (\bm{3},\bm{2},-5/6)_i \oplus
(\overline{\bm{3}},\bm{2},5/6)_i \oplus (\bm{1},\bm{1},0)_i= \rho^i_{8} \oplus \rho^i_{3} \oplus \rho^i_{(3,2)} \oplus \rho^i_{(\bar{3}, 2)} \oplus \rho^i_{24}$, $i=1,2$ --- in order to generate experimentally viable neutrino masses and mixing parameters. This extension can accommodate two massive neutrinos that are sufficient, at this stage, to reproduce the observed values of squared mass differences. 

The $SU(5)$ invariant contractions that are relevant for the mass generation of the SM gauge group multiplets in $\bm{24}_i$ are $m_{i j} \bm{24}_{i \beta}^\alpha \bm{24}_{j \alpha}^\beta$ and $\lambda_{i j} \bm{24}_{i \beta}^\alpha \bm{24}_{j \gamma}^\beta \bm{24}_{\alpha}^\gamma$, where $i,j=1,2$ and $\alpha,\beta,\gamma=1,\ldots,5$. 

Let us, for a moment, consider a scenario with only one adjoint representation --- $\bm{24}_1$ --- to see how many free parameters that describe the mass splitting we have at our disposal. One adjoint has four multiplets under the SM gauge group. The component that transforms as $(\bm{1},\bm{1},0)$ is not relevant for a discussion of the gauge coupling unification and we disregard it in what follows. We also note that the components that transform as $(\bm{3},\bm{2},-5/6)$ and $(\overline{\bm{3}},\bm{2},5/6)$ are degenerate in mass. This then means that there are three relevant mass scales, i.e., $m_{\rho^1_8}$, $m_{\rho^1_3}$ and $m_{\rho^1_{(3,2)}} \equiv m_{\rho^1_{(\bar{3}, 2)}}$, to be considered. We have, on the other hand, two independent $SU(5)$ contractions that contribute towards these masses. We thus need to specify two mass scales to fix the third one. Simply put, if $m_{\rho^1_3}$ and $m_{\rho^1_8}$ are given, one can evaluate $m_{\rho^1_{(3,2)}} \equiv m_{\rho^1_{(\bar{3}, 2)}}$. Indeed, the relevant mass relations are $
m_{\rho^1_8}=\hat{m}_1 m_{\rho^1_3}$, $m_{\rho^1_{(3,2)}} \equiv m_{\rho^1_{(\bar{3}, 2)}}=(1+\hat{m}_1)m_{\rho^1_3}/2$, where $\hat{m}_1$ is a free parameter~\cite{Perez:2007rm} that conveniently describes the mass splitting between $m_{\rho^1_3}$ and $m_{\rho^1_8}$.

It is now straightforward to apply the preceding discussion on the scenario with two adjoint representations. If the matrices with matrix elements $m_{i j}$ and $\lambda_{i j}$, $i,j=1,2$, are taken to be diagonal we would have two sets of equations for the SM gauge group multiplets in $\bm{24}_i$, $i=1,2$, that would read
\begin{align}
\label{eq:jjjj1}
m_{\rho^i_8}&=\hat{m}_i m_{\rho^i_3},\\
\label{eq:jjjj2}
m_{\rho^i_{(3,2)}}&\equiv m_{\rho^i_{(\bar{3}, 2)}}=\frac{(1+\hat{m}_i )}{2}m_{\rho^i_3},
\end{align}
where $\hat{m}_i$ are free parameters. To simplify our analysis of gauge coupling unification we take $\hat{m}_1=\hat{m}_2 \equiv \hat{m}$ in what follows. It can be explicitly demonstrated that this assumption does not influence the search for the maximal possible value of $m_\mathrm{GUT}$. We furthermore place the following constraints on the masses of the fields that can affect the running of the gauge coupling constants: $3 \times 10^{11}\,\mathrm{GeV} < m_{\Psi_T} \leq m_\mathrm{GUT}$~\cite{Dorsner:2012uz}, $10^{5}\,\mathrm{GeV} < m_{\Sigma_8} \leq m_\mathrm{GUT}$~\cite{Bajc:2006ia} and $4 \times 10^2\,\mathrm{GeV} \leq m_{\rho^i_3}, m_{\rho^i_8}, m_{\rho^i_{(3,2)}}, m_{\Sigma_3}  \leq m_\mathrm{GUT}$. We also demand that Eqs.~\eqref{eq:jjjj1} and~\eqref{eq:jjjj2} hold.

Again, we consider two cases. This time we start with the scenario when $m_4^D=800$\,GeV and demand that $m_4^E>1.2$\,TeV. The maximal value of the GUT scale is $m_\mathrm{GUT}=10^{15.72}$\,GeV and it is achieved for the following values of masses: $m_{\Sigma_3}=10^{15.72}$\,GeV, $m_{\Sigma_8}=10^{5}$\,GeV,  $m_{\Psi_T}= 3 \times 10^{11}$\,GeV,  $m_{\rho^1_8}=m_{\rho^2_8}=10^{15.72}$\,GeV,  $m_{\rho^1_3}=m_{\rho^2_3}=10^{5.35}$\,GeV,  $m_{\rho^1_{(3,2)}}=m_{\rho^2_{(3,2)}}=10^{15.42}$\,GeV, $m_4^D=800$\,GeV and $m_4^E=10^{15.72}$\,GeV. One can see that this solution corresponds to $\hat{m}=10^{10.37}$. There exists no possibility for gauge coupling unification for $\hat{m}>10^{10.37}$. Note, however, that the value of $\hat{m}$ can be decreased down to $10^{8.73}$ without any change in the value of $m_\mathrm{GUT}$. Namely, this only affects the splitting between the masses of $\rho^i_3$ and $\rho^i_8$, where $i=1,2$. Once $\hat{m}$ is below $10^{8.73}$, the GUT scale starts to slowly decrease. (There exists no unification for $\hat{m} < 10^{2.37}$. The end point $\hat{m} = 10^{2.37}$ yields $m_\mathrm{GUT}=10^{14.29}$\,GeV.)

The proposed scenario is extremely predictive, since the upper bound on the GUT scale is $m_\mathrm{GUT} \leq 5.3 \times 10^{15}$\,GeV. The associated value of the inverse of the gauge coupling at the GUT scale is found to be $\alpha_\mathrm{GUT}^{-1}=35.0$. One can see from Fig.~\ref{fig:p_2} that a part of the available parameter space is already excluded by the current experimental limit on $p \rightarrow \pi^0 e^+$ in the case when $m_4^D=800$\,GeV in the model that incorporates the vector-like pair $(\bm{5}_4,\overline{\bm{5}}_4)$ of matter fields and two $24$-dimensional representations of fermions. The upper bound on $m_\mathrm{GUT}$ is shown as the dashed horizontal line in Fig.~\ref{fig:p_2}. Note that the lower limit on $m_\mathrm{GUT}$ that is shown in Fig.~\ref{fig:p_2} for the $m_4^D=800$\,GeV case is evaluated for $\alpha_\mathrm{GUT}^{-1}=35.0$ to make the comparison self-consistent. We stress that this model accommodates all experimentally observed fermion masses and mixing parameters. It is also in agreement with all relevant experimental constraints within the allowed region of parameters space when $m_1 > 0.055$\,GeV. (See Fig.~\ref{fig:p_2}.)

The second case, when $m_4^E=400$\,GeV, cannot be accommodated within this extension. Namely, the maximal value of the GUT scale is below the bound inferred from proton decay experiments. Namely, we find that $m_\mathrm{GUT}=10^{14.72}$\,GeV for $m_4^E=400$\,GeV.


\section{Conclusions}
\label{CONCLUSIONS}

We study the original Georgi-Glashow $SU(5)$ model extended with one vector-like down-type quark and one vector-like lepton doublet. These comprise a pair of five-dimensional representations of $SU(5)$ and help to obtain correct values of the down-type quark and charged lepton masses.

All unitary transformations that relate an arbitrary flavor basis to the mass eigenstate basis of matter fields are completely described with three parameters in this scenario. These, on the other hand, are limited to reside in very narrow ranges due to existing experimental data. We accordingly find a clear correlation between these parameters, proton decay and the lightness of either vector-like quarks or vector-like leptons, with the light vector-like lepton case being more restrictive than the light vector-like quark case. Representative mass scales for vector-like quarks and leptons are taken to be $800$\,GeV and $400$\,GeV, respectively.

We investigate the viability of the scenario when either the vector-like quark or the vector-like lepton states are light, taking into account relevant low-energy constraints. These, for example, include the influence of vector-like leptons on $\mu$--$e$ conversion and the modification of the couplings of the SM fermions to the $Z$ boson due to the presence of the vector-like state. Our study demonstrates that the proposed framework is very predictive with regard to proton decay signatures through gauge boson mediation. We find the most relevant decay channels to be $p \rightarrow \pi^0 e^+$ and $p \rightarrow \pi^0 \mu^+$ when vector-like leptons are taken to be $400$\,GeV. The associated lower bound on the GUT scale turns out to be between $4.2 \times 10^{15}$\,GeV and $5.7 \times 10^{15}$\,GeV for $\alpha_\mathrm{GUT}=0.033$. A lower bound on the GUT scale when vector-like quarks are assumed to be $800$\,GeV is generated solely by $p \rightarrow \pi^0 e^+$. It goes from $3.5 \times 10^{15}$\,GeV to $5.4 \times 10^{15}$\,GeV for $\alpha_\mathrm{GUT}=0.029$.

We discuss two realistic extensions of the minimal $SU(5)$ model with vector-like fermions that do not interfere with the predictions from the low-energy constraints that we implement. The first extension is based on an additional $50$-dimensional scalar representation. It yields a viable gauge coupling unification and allows for the presence of a light vector-like down-type quark and/or a vector-like lepton doublet. The second scenario includes two additional adjoint representations of fermions. It provides a viable gauge coupling unification only when vector-like down-type quark states are light. It yields an upper bound on the GUT scale of $5.3 \times 10^{15}$\,GeV with $\alpha_\mathrm{GUT}=0.029$ when the mass of the vector-like down-type quark states is taken to be $800$\,GeV. 


\begin{acknowledgments}
I.D.\ acknowledges the SNSF support through the SCOPES project No.\ IZ74Z0\_137346. The work of S.F.\ is supported by ARRS. I.M.\ acknowledges the support by the Ad futura Programme of the Slovenian Human Resources and Scholarship Fund and ARRS. We thank J.F.\ Kamenik and A.\ Greljo for insightful discussions.
\end{acknowledgments}


\end{document}